\documentclass[prl,aps,epsfig,showpacs,superscriptaddress]{revtex4}
\usepackage{bm}
\usepackage{amsfonts}
\usepackage[dvips]{graphicx}
\usepackage{mathrsfs}
\usepackage[intlimits]{amsmath}
\usepackage[colorlinks, citecolor=red]{hyperref}

\begin{document}

\title{Room-temperature storage of quantum entanglement using
decoherence-free subspace in a solid-state spin system}
\author{F. Wang$^1$, Y.-Y. Huang$^1$, Z.-Y. Zhang$^{1,2}$, C. Zu$^1$, P.-Y. Hou$^1$, X.-X. Yuan$^1$, W.-B.
Wang$^1$, W.-G. Zhang$^1$, L. He$^1$, X.-Y. Chang$^1$, L.-M. Duan}
\affiliation{Center for Quantum Information, IIIS, Tsinghua University, Beijing 100084,
PR China}
\affiliation{Department of Physics, University of Michigan, Ann Arbor, Michigan 48109, USA}
\date{\today }

\begin{abstract}
We experimentally demonstrate room-temperature storage of quantum
entanglement using two nuclear spins weakly coupled to the electronic spin
carried by a single nitrogen-vacancy center in diamond. We realize universal
quantum gate control over the three-qubit spin system and produce entangled
states in the decoherence-free subspace of the two nuclear
spins. By injecting arbitrary collective noise, we demonstrate that the
decoherence-free entangled state has coherence time longer than that of
other entangled states by an order of magnitude in our experiment.
\end{abstract}

\maketitle

\section{Introduction}

Decoherence caused by the system-environment interaction poses a serious
obstacle to physical implementation of quantum information processing \cite%
{1,2}. Strategies involving active interventions, such as dynamical
decoupling \cite{3,4,5,6,7,8,9,10} and quantum error correction \cite%
{11,12,13,14}, have been extensively studied in experiments to recover
quantum information from coupling with the environment \cite{15,16,17,18}.
Meanwhile, passive error control methods with no active recovery have also
been proved to be efficient in preventing collective decoherence caused by
symmetric system-environment coupling \cite{19,20,21,22,23,24,25,26}.
Quantum information in the decoherence-free subspace (DFS) does not
decohere and is well protected even with perturbation in the
system-environment interaction, making DFS an ideal quantum memory. DFS has
been demonstrated in several experimental systems to protect single qubits
from collective dephasing \cite{27,28,29,30,36}.

In this paper, we present an experimental demonstration of DFS in a
room-temperature solid-state system and use DFS to store quantum
entanglement against general collective noise including both dephasing and
dissipation. Quantum storage of single qubits has been demonstrated in a
number of experimental systems, including trapped ions \cite{36a}, single
nuclear spins \cite{36}, atomic or spin ensembles \cite{36b,36c,36d}. To realize
the full capability of quantum memory, it is important to further extend the
information storage from single qubits to quantum entanglement. This
extension is not straightforward as the best quantum memories demonstrated
so far typically require good isolation of the qubits, which makes it
difficult to generate entanglement between the qubits in the same system. Entanglement between nuclear spins coupled to the NV centers have been created in multiple works \cite{17,34,36,38a,38b,38c}.
Here we extend these works by demonstrating room-temperature storage of quantum entanglement in the DFS with
two nuclear spins and the effectiveness of DFS under general 
collective noise. We produce entanglement between the
nuclear spins within the DFS through universal gate control on the electronic and the nuclear spins. 
Under general collective noise, we demonstrate that
the entangled state in DFS has coherence time longer than that of other
entangled states by an order of magnitude.

\section{Results}

\subsection{Decoherence-free subspace}

A DFS takes advantage of qubit-permutation symmetry in the
system-environment interaction to isolate the stored quantum information
from the environment. Therefore, evolution of quantum states inside a DFS is
purely unitary. A simple example for a DFS is provided by the two-qubit
subspace spanned by $|0\rangle _{D}=|0\rangle _{n1}|1\rangle _{n2}$ and $|1\rangle
_{D}=|1\rangle _{n1}|0\rangle _{n2}$ when these two qubits are subject to
collective dephasing noise \cite{19,20}. Apparently, a collective random
phase $\phi $ accumulated for the basis states $|0\rangle \rightarrow
e^{i\phi }|0\rangle ,|1\rangle \rightarrow e^{-i\phi }|1\rangle $ cancel out
in this subspace. Most of the experimental demonstrations focus on this
special case \cite{27,28}. Under general collective noise including both
dephasing and relaxation, the states $|0\rangle _{D}$ and $|1\rangle _{D}$
are not stable any more, but their combination, the singlet state $|S\rangle
=(|0\rangle _{n1}|1\rangle _{n2}-|1\rangle _{n1}|0\rangle _{n2})/\sqrt{2}$
is still an entangled state lying within the DFS \cite{22,23,24}.

\subsection{Control of two weakly coupled nuclear spins}

We use two $C^{13}$ nuclear spins weakly coupled to an individual NV center
electronic spin in a diamond crystal as our qubits (Fig. 1(a,b)). The NV electronic spin
is a well characterized spin-$1$ system which can be optically initialized
and readout \cite{31}, and coherently manipulated with microwave source at
room temperature \cite{32}. We use the NV electronic spin as a handle to
coherently control and entangle the nuclear spins and read out their final
state \cite{33,34,35}. The external magnetic field provides a source of
collective dephasing noise to the target nuclear spins. We prepare two
typical entangled states $|T\rangle =(|0\rangle _{n1}|1\rangle
_{n2}+|1\rangle _{n1}|0\rangle _{n2})/\sqrt{2}$ and $|S\rangle =(|0\rangle
_{n1}|1\rangle _{n2}-|1\rangle _{n1}|0\rangle _{n2})/\sqrt{2}$ to
demonstrate the DFS under the collective dephasing noise and find that the
memory time is limited by the electronic spin relaxation time $T_{1}$. To
verify the DFS under arbitrary collective noise including both dephasing and
relaxation, we realize a general collective noise model by injecting a noisy
radio frequency field into the system \cite{34,37}. Under general collective
noise, we show that the entangled state $|S\rangle $ within the DFS is still
well protected until the electronic spin relaxation breaks the
system-environment symmetry while the state $|T\rangle $ quickly decoheres.

\begin{figure}[tbp]
\includegraphics[width=180mm]{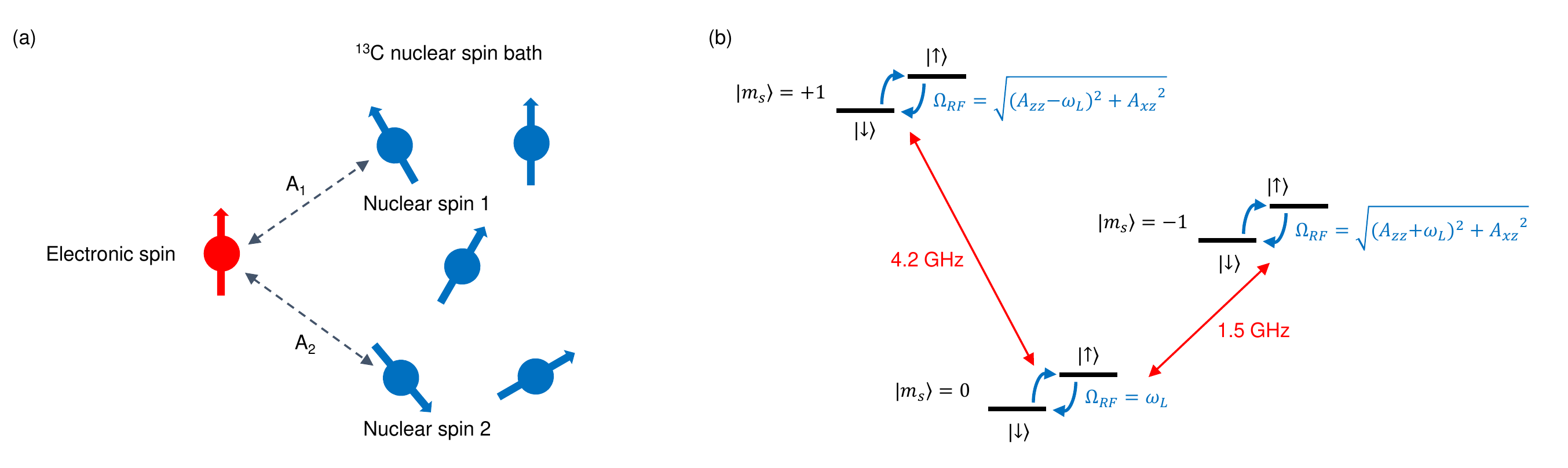}
\caption{Experimental system. (a) The NV electronic spin (red) and the coupled $^{13}C$ spin bath (Blue). Entanglement states are stored in two isolated weakly coupled $^{13}C$ nuclear spins. (b) Energy structure of the NV electronic spin and a weakly coupled nuclear spin. Nuclear spin sublevels $|\uparrow\rangle$ and $|\downarrow\rangle$ are split by Zeeman shift ($\omega_L$) and hyperfine interaction ($A_{zz}$, $A_{xz}$) with $\omega_0=\omega_L(m_s=0)$, $\omega_{\pm 1}=\sqrt{(A_{zz}\mp\omega_L)^2+A_{xz}^2}(m_s=\pm 1)$}
\end{figure}

\begin{figure}[tbp]
\includegraphics[width=180mm]{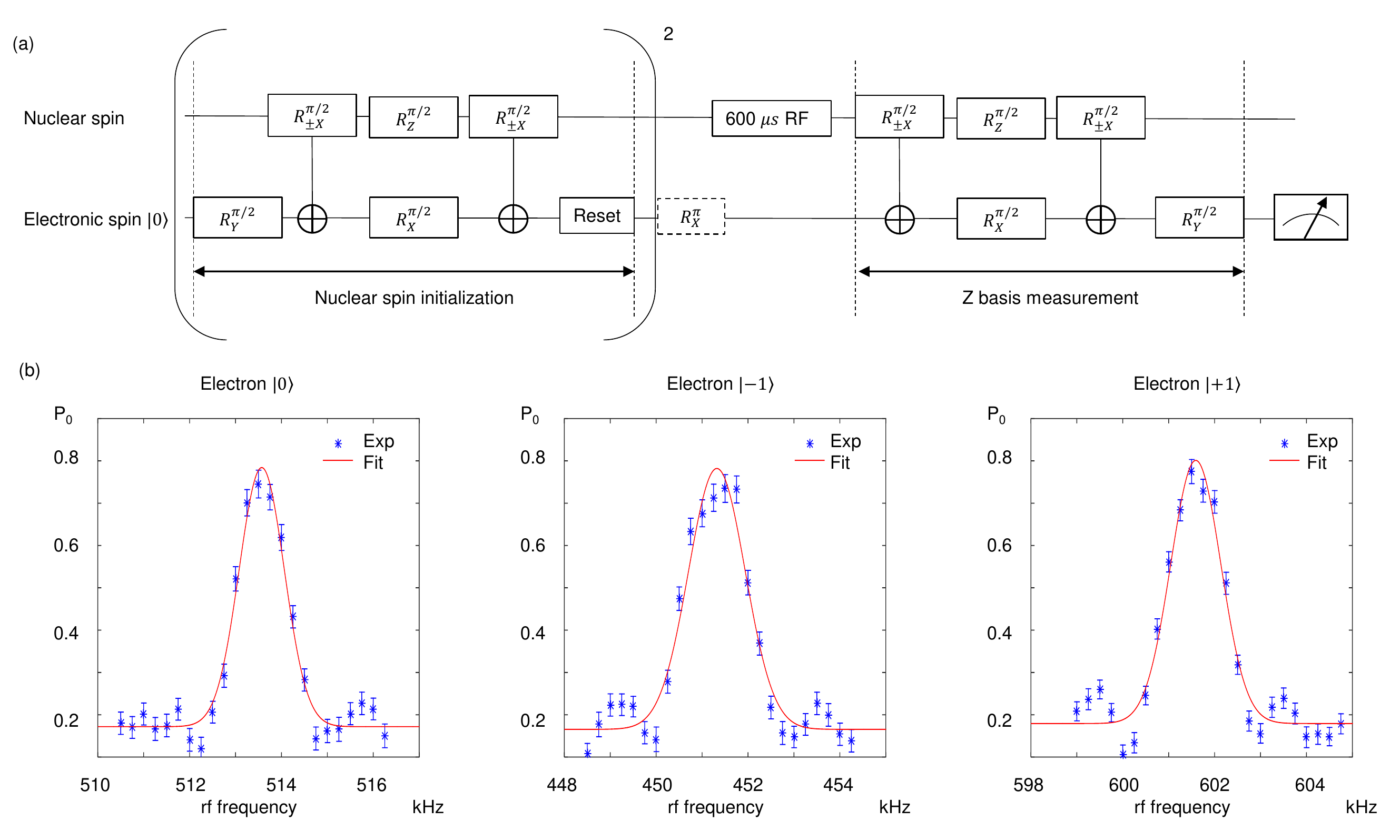}
\caption{Calibration of the nuclear spin hyperfine interaction parameters. (a) Gate sequence to scan
the resonant frequency of nuclear spins with the electronic spin set at $m_s=+1,0,-1$ states, respectively. The nuclear spin is initialized by swapping the electronic spin polarization onto the nuclear spin. Electronic spin is reset to $|0\rangle$ or $|\pm 1\rangle$ state using a $350$ ns green laser or an additional $\protect%
\pi$ rotation afterwards. A rf pulse with a duration of $600$ $\mu$s and a
scanning frequency is then implemented on the nuclear spin to trigger spin
flips at resonant frequency. The final state readout is accomplished by
swapping the nuclear spin state back onto the electronic spin. (b) Probability of electronic spin in $m_s=0$ state ($P_0$) as a function of the rf frequency
with the electronic spin at $m_s=0,-1,+1$ states, respectively, for nuclear spin 1. Solid lines are the Gaussian fits. See the supplementary material for results on the nuclear spin 2.}
\end{figure}

The experiments are performed at room temperature on a diamond sample with
an external magnetic field of $480$ Gauss along the NV symmetry axis. We use
the hyperfine interaction to coherently manipulate the nuclear spin by
applying an equally-spaced sequence of $\pi $ rotations (the
Carr-Purcell-Meiboom-Gill, or CPMG sequence) to flip the electronic spin
\cite{17,18,35}. We use the XY8 sequence in our experiment to reduce the
influence of imperfection in pulse durations and the accumulation of
systematic pulse errors \cite{16,37}. The multi-pulse CPMG\ sequence
decouples the electronic spin from the spin bath. At the same time, the
electronic spin gets entangled with a specific nuclear spin when the pulse
interval $2\tau $ satisfies certain resonance condition, which leads to
collapse of the electronic spin coherence after the CPMG\ sequence and thus
can be detected. The resonance condition depends on $A_{\parallel }$, the
parallel component of the hyperfine interaction for the specific nuclear spin,
and is given by
\begin{equation*}
2\tau \approx \frac{2(2k-1)\pi }{2\omega _{L}+A_{\parallel }}
\end{equation*}%
where the integer $k$ denotes the order of resonance and $\omega _{L}$ is the nuclear spin
Larmer frequency. Based on this resonance, we control the total number of $%
\pi $ pulses $N$ and the pulse interval $2\tau $ to complete single-bit
operations (X or Z rotation) or conditional operation ($\pm $X rotation
conditional on the state of electronic spin) on the target nuclear spins,
where X and Z denote the Pauli matrices $\sigma _{x}$ and $\sigma _{z}$. For
each type of gates, the condition for $N$ depends on the transverse component of
the hyperfine interaction $A_{\perp }$ \cite{39}.

\begin{figure}[tbp]
\includegraphics[width=180mm]{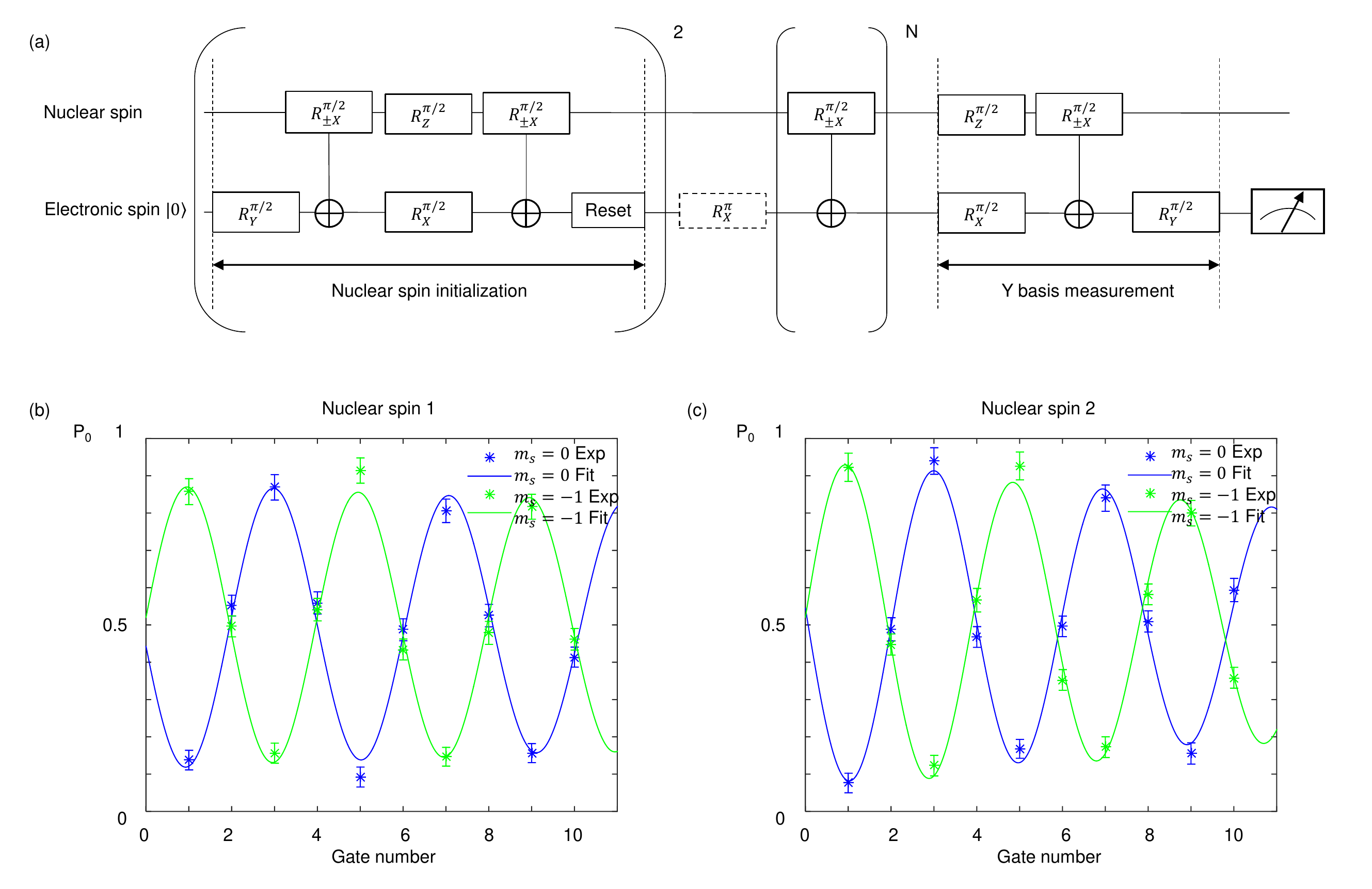}
\caption{Characterization of the conditional X gate on nuclear spins 1 and
2. See the supplementary material for results on unconditional gates. (a)
Experimental scheme to characterize the gate fidelity. The nuclear spin is
polarized by swapping the electronic spin polarization onto the nuclear
spin. An additional $\protect\pi$ rotation is applied to set the electronic
spin to $m_s=-1$ state. After that, the desired gate (conditional X gate) is
applied on the nuclear spin for $N$ times ($N=1,...,10$) with the electronic spin
at $m_s=0$ or $m_s=-1$ before measuring the nuclear spin on the Y
basis. (b,c) Experimental results of conditional X gate on the nuclear spin 1 and
2, respectively. The nuclear spin rotates on the opposite direction of the X axis with
the electronic spin at $m_s=0,-1$ states. Solid lines are fits by the function $sin(2%
\protect\pi N/4)(1-bN)$ with
$b=0.012$ and a standard deviation of $\sigma=0.011$ in (b) and $b=0.025$ and a standard deviation of $\sigma=0.014$ in
(c). The results are without correction of initialization and detection error.}
\end{figure}

\subsection{Calibration of hyperfine parameters}

To perform high-fidelity gate operations on the weakly coupled nuclear
spins, it is required to have precise calibration of the hyperfine
interaction magnitudes $A_{\parallel }$ and $A_{\perp }$ for each target
nuclear spins. The hyperfine parameters can be calibrated with a resolution
about $10$ kHz by fitting the experimental data on the measured electronic
spin coherence after the CPMG\ sequence to the numerical simulation of the
corresponding dynamics with the fitting parameters $A_{\parallel }$ and $%
A_{\perp }$. However, as the gate fidelity is strongly correlated with the
precision of the hyperfine parameters, the $10$ kHz resolution in
calibration is not enough for achieving high-fidelity quantum gates on the
nuclear spins. We describe a method based on the nuclear spin ODMR (Optical
Detected Magnetic Resonance) for high-precision calibration of $A_{\parallel
}$ and $A_{\perp }$ in experiments. We measure the resonant frequency of the
nuclear spins with the electronic spin set at $m_{s}=+1,0,-1$ respectively.
As described in Fig. 2(a), with rough calibration of the hyperfine
parameters by the CPMG\ sequence, we first polarize the nuclear spin (with
significant imperfection) by swapping the electronic spin polarization onto
the nuclear spin, and optically reset the electronic spin to $m_{s}=0$ state
(or $m_{s}=\pm 1$ state by another resonant microwave $\pi $ rotation).
After that, we apply a $\pi $-pulse of $600$ $\mu s$ duration on the target
nuclear spin using radio frequency field and measure the nuclear spin flip
probability by swapping the nuclear spin polarization back onto the
electronic spin. In Fig. 2(b), we show that this approach gives a
resonant frequency with a standard deviation of $0.05$ kHz, thus allows us
to determine the nuclear spin hyperfine parameter to a resolution about $0.05
$ kHz in the parallel component $A_{\parallel }$ and about $0.5$ kHz in the
transverse component $A_{\perp }$ \cite{39}.

After the hyperfine parameters are precisely calibrated, we perform the
desired gate (conditional X gate, unconditional X and Z gate) on the
polarized nuclear spins with electronic spin at $m_{s}=0$ or $m_{s}=-1$
state. To estimate the gate fidelity, we apply the same gate $10$ times, and
from the slow decay of the target state fidelity as shown in Fig. 3 and the
supplementary material, we extract a gate fidelity about $F\approx 0.988$ ($F\approx 0.975$) for the conditional operations on nuclear spin 1 (spin 2). Gate fidelity for nuclear spin 1 is slightly higher than that for nuclear spin 2, because nuclear spin 1 has a larger parallel component of hyperfine parameters, which leads to a shorter gate time \cite{39}. Using the high fidelity
conditional X gate and the unconditional Z gate, single nuclear spin
initialization and readout fidelity is enhanced to $F_{1}=0.896(6)$ and $%
F_{2}=0.873(9)$ for nuclear spin 1 and 2 \cite{39}.

\begin{figure}[tbp]
\includegraphics[width=180mm]{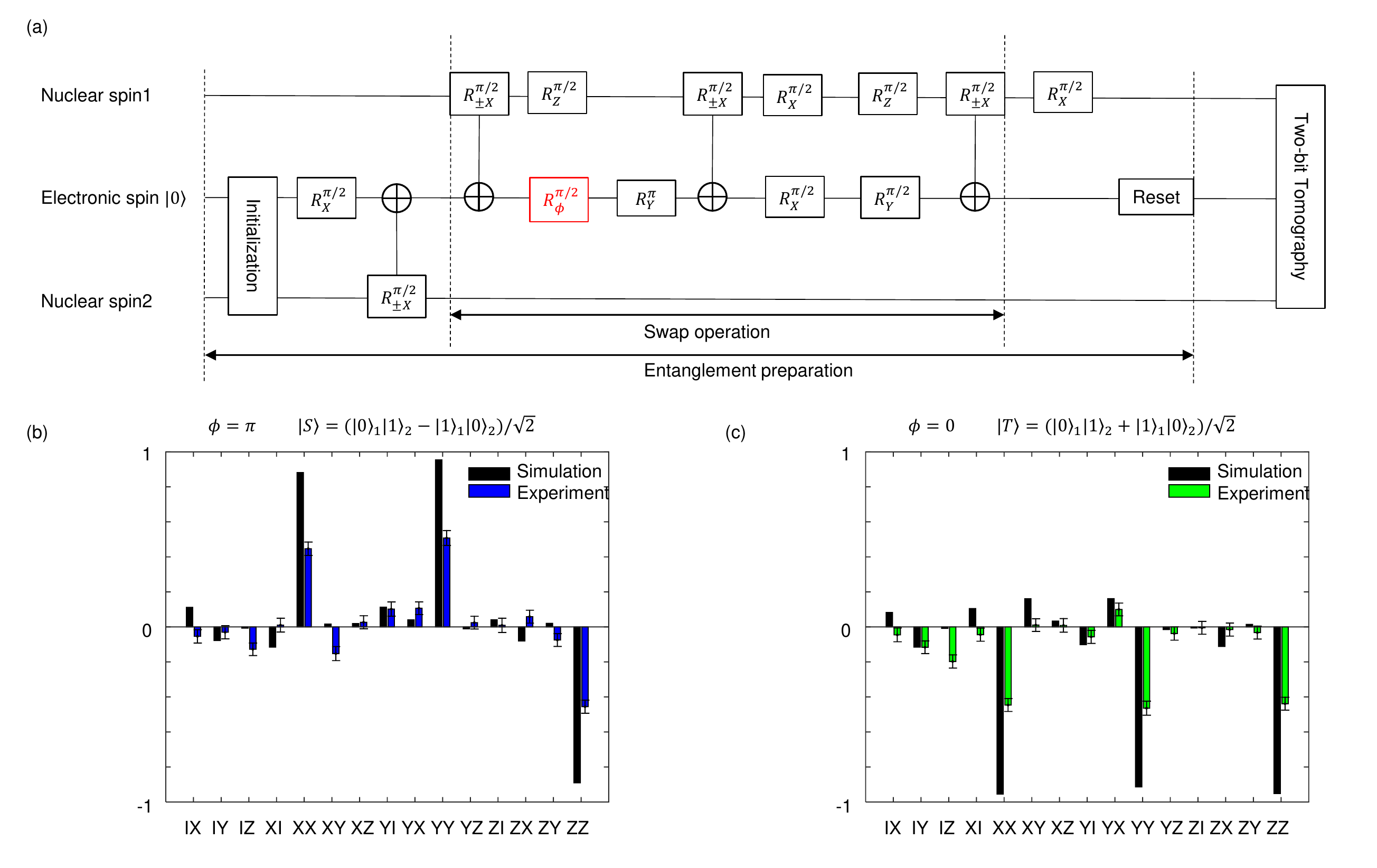}
\caption{Preparation and detection of entangled states between the nuclear spins. (a)
Gate sequence to prepare entangled states between nuclear spins at room temperature.
Entanglement is first generated between the nuclear spin 2 and the electronic
spin. By swapping the electronic spin with the nuclear spin 1,
entanglement between nuclear spins is produced. A subsequent $\protect\pi/2$ rotation is applied
to prepare the entangled state in the DFS. The phase $\protect\phi$ of the
operation in red is controlled to produce $|T\rangle$ state ($\protect\phi=0$%
) or $|S\rangle$ state ($\protect\phi=\protect\pi$). The readout is
performed by quantum state tomography on the two nuclear spins \protect\cite{39}.
(b,c) Quantum state tomography results for $|S\rangle$ state (b) and $|T\rangle$
state (c). Black bar describes the simulation result \protect\cite{39}, blue (green)
bar is the experiment data for $|S\rangle$ ($|T\rangle$) state without correction of
initialization and readout errors. }
\end{figure}

\subsection{Entanglement preparation}

We prepare two typical entangled states $|T\rangle $ and $|S\rangle $ for
the nuclear spins using the above gates. When the nuclear spins are subject
to collective dephasing noise, both the two states are decoherence free.
However, only the entangled state $|S\rangle $ is protected under arbitrary
collective noise. To produce the desired entangled states, as shown in Fig.
4(a), we first prepare an electron-nuclear entangled state $(|0\rangle
_{e}|Y_{-}\rangle _{n2}-i|1\rangle _{e}|Y_{+}\rangle _{n2})/\sqrt{2}$ by
applying a conditional $\pi /2$ operation on the polarized nuclear spin 2
with the electronic spin set at $(|0\rangle -i|1\rangle )/\sqrt{2}$ state,
where $|Y_{\pm }\rangle $ denotes the eigenstate of $\sigma _{y}$ with $\pm 1
$ eigenvalue. After that, we coherently swap the states between the
electronic spin and the nuclear spin 1 by applying a sequence of gate
operations as shown in Fig. 4(a), and subsequently implement a single-bit X
gate on the nuclear spin 1 to produce the target entangled states within the
DFS of the two nuclear spins. By controlling the phase $\phi $ of the swap
gate we are able to prepare the entangled state to either $|T\rangle $ or $%
|S\rangle $. The entangled state fidelity is characterized by calculating
the overlap between the experiment density matrix $\rho _{exp}$ constructed
through quantum state tomography \cite{39} and the target ideal state $|\Psi
_{id}\rangle $ through $F=\left\langle \Psi _{id}\right\vert \rho
_{exp}|\Psi _{id}\rangle $. With the measured fidelity $F=0.60(1)$ for $%
|S\rangle $ state and $F=0.59(1)$ for $|T\rangle $ state (without correction
of initialization and detection error), we demonstrate entanglement between
the nuclear spins (Fig. 4(b,c)).

Various imperfections affect the entangling process, which leads to a low
entangled state fidelity. We summarize the four major contributions. (i) The
preparation process involves the initialization of nuclear spin 2, with a
single-qubit initialization and readout fidelity about $0.87$, we expect a
similar fidelity drop in term of the entanglement fidelity. (ii) The use of
green laser at the end of the entangling process to optically reset the
electronic spin decreases the nuclear spin fidelity in both polarization and
coherence \cite{17,38}. (iii) The intrinsic errors mostly caused by the
crosstalk between the targeted two nuclear spins decrease the entangled
state fidelity from $1$ to $0.95$ in our numerical simulation (see Fig.
4(b,c)). (iv) Decoherence, magnetic field fluctuation and gate error
accumulation in each experimental run (note that the whole state preparation
process requires application of more than ten gates) reduce the final state
fidelity over the $10^{6}$ repetitions of experiments for measurement of
each density matrix element \cite{39}. At room temperature, due to these limitations, 
it is hard to significantly improve the entanglement fidelity for the nuclear spins. 
With an isotopically purified samples, the coherence time for the electronic spin increases, but it
becomes more difficult to find nuclear spins with appropriate hyperfine interaction strength
for the entangling gates. If we put the sample in a cryogenic environment, both the initialization 
fidelity and the coherence time for the electronic spin would be significantly improved,
and correspondingly the entanglement fidelity for the nuclear spins will increase substantially \cite{36}.  

\begin{figure}[tbp]
\includegraphics[width=180mm]{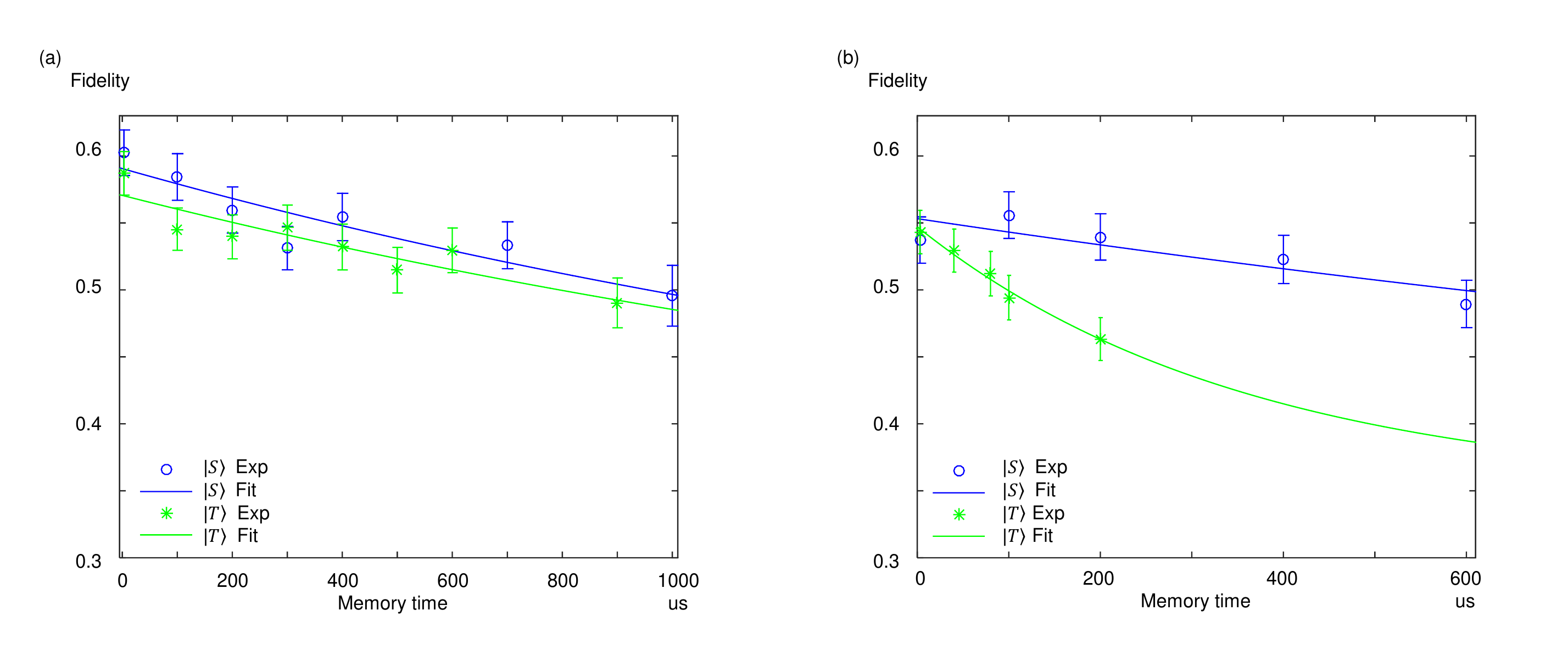}
\caption{Decay of the entanglement fidelity under various noise environments.
(a) Entanglement fidelity as a function the storage time $t$ under collective
dephasing noise. Solid lines are fits to $exp(-t/T_{est})$ with $%
T_{est}=2.24$ ms and a standard deviation of $\sigma=153$ $\mu$s for $|S\rangle$ state (blue) and $T_{est}=2.29$ ms and a standard deviation of $\sigma=232$ $\mu$s for $%
|T\rangle$ state (green). The fitting curves saturate at $0.35$, which
corresponds to the fidelity of the final state when all the coherence terms drop to zero. Due to the limited
fidelity for initial state preparation, the population is not given by an identity matrix, so the saturation fidelity is
$0.35$ instead of $0.5$ . (b) Entanglement fidelity as a function of the storage
time $t$ under general collective noise. Solid lines are fits to $%
exp(-t/T_{est})$ with $T_{est}=2.18$ ms and a standard deviation of $\sigma=366$ $\mu$s for $|S\rangle$ state (blue) and $%
T_{est}=360$ $\mu$s and a standard deviation of $\sigma=20$ $\mu$s for $|T\rangle$ state (green).}
\end{figure}

\subsection{Test of DFS under collective dephasing noise}

We start by exploring DFS with the system subject to a collective dephasing
noise, which in our case is the external magnetic field. In Fig 5(a), we
prepare the nuclear spins in the DFS and measure their state fidelity extracted
from quantum state tomography as a function of storage time. By fitting the
data to $exp\left( -t/T_{est}\right) $, we extracted a memory time of $%
T_{est}\approx 2.3$ $ms$, which is limited by the electronic spin relaxation time
$T_{1}\approx 2.5$ $ms$. This can be explained by the breakup of the
system-environment coupling symmetry. As the electronic spin relaxes, it
causes independent dephasing noise for the two nuclear spins with $\Delta
\omega \approx \left\vert A_{\parallel 1}-A_{\parallel 2}\right\vert \approx
148$ $kHz$, which destroys the state quickly \cite{36}. Longer memory time
could be achieved for entangled states if one makes use of the isotopically
purified diamond samples to reduce the nuclear spin crosstalk error with spin bath and
repeatedly polarizes the electronic spin to mitigate the dephasing noise
\cite{38}. Alternatively, if one put the diamond sample in the cryogenic
environment, both the entanglement fidelity and entanglement storage time
can be significantly improved as the electronic spin relaxation time gets
much longer under low temperature \cite{38d}.

\subsection{Test of DFS under general collective noise}

A crucial step to verify DFS is to investigate the state coherence under
general collective noise including both dephasing and relaxation. To realize
general collective noise in addition to the dephasing induced by the
external magnetic field, we introduce collective relaxation by injecting a
noisy radio-frequency field. Because the magnetic field couples the nuclear
spins identically, the relaxation induced by the injected rf field is
collective to nuclear spins in the close neighborhood of the electronic
spin. In Fig. 5(b), We compare the storage time of two typical entangled
states $|T\rangle $ and $|S\rangle $. In agreement with theory, only $%
|S\rangle $ state which lies within the DFS under arbitrary collective
noise is protected against the injected noise with a fitted memory time $%
T_{est}\approx 2.2$ $ms$. In comparison, $|T\rangle $ state is destroyed
quickly with a fitted memory time $T_{est}\approx 360$ $\mu s$.

\section{Summary}

In summary, we have demonstrated room temperature storage of quantum
entanglement by preparing quantum states in the DFS of two nuclear spins
and experimentally verified that the entangled state within the DFS\ has
coherence time significantly longer than that of other components under
general collective noise. Storage of quantum entanglement is required in
many quantum information protocols and our result suggests that the DFS
could find interesting applications in experimental realization of those
protocols.

We thank T. H. Taminiau for discussions. This work was supported by Tsinghua University
and the Ministry of Education of China. LMD and ZYZ acknowledge in addition support from the AFOSR MURI and the
ARL CDQI program.


\begin{thebibliography}{99}
\bibitem{1} W. G. Unruh, Phys. Rev. A 51, 992 (1995).

\bibitem{2} D. P. DiVincenzo, Fortschritte der Physik 48, p. 771 (2000).

\bibitem{3} L. Viola and S. Lloyd, Phys. Rev. A 58, 2733 (1998).

\bibitem{4} P. Zanardi, Phys. Lett. A 258, 77 (1999).

\bibitem{5} L.-M. Duan and G. Guo, Phys. Lett. A 261, 139 (1999).

\bibitem{6} D. Vitali and P. Tombesi, Phys. Rev. A 59, 4178 (1999).

\bibitem{7} L. Viola, E. Knill, and S. Lloyd, Phys. Rev. Lett. 82, 2417
(1999).

\bibitem{8} L. Viola and E. Knill, Phys. Rev. Lett. 90, 037901 (2003).

\bibitem{9} K. Khodjasteh and D. A. Lidar, Phys. Rev. Lett. 95, 180501
(2005).

\bibitem{10} H. K. Ng, D. A. Lidar, J. Preskill, Phys. Rev. A 84, 012305
(2011).

\bibitem{11} P. W. Shor, Phys. Rev. A 52, R2493 (1995).

\bibitem{12} A. M Steane, Phys. Rev. Lett. 77, 793 (1996).

\bibitem{13} A. R. Calderbank and P. W. Shor, Phys. Rev. A 54, 1098 (1996).

\bibitem{14} A. M. Steane, Proc. R. Soc. London A 452, 2551 (1996).

\bibitem{15} G. de Lange, Z. H. Wang, D. Riste, V. V. Dobrovitski, R.
Hanson, Science 330, 60 (2010).

\bibitem{16} C. A. Ryan, J. S. Hodges, and D. G. Cory, Phys. Rev. Lett. 105,
200402 (2010).

\bibitem{17} T. H. Taminiau, J. Cramer, T. van der Sar, V. V. Dobrovitski
and R. Hanson, Nature Nanotech. 9, 171-176 (2014).

\bibitem{18} J. Cramer, N. Kalb, M. A. Rol, B. Hensen, M. S. Blok, M.
Markham, D. J. Twitchen, R. Hanson and T. H. Taminiau, Nat. Commun. 7: 11526
(2016).

\bibitem{19} G. M. Palma, K.-A. Suominen and A. K. Ekert, Proc. Roy. Soc.
London Ser. A 452: 567 (1996).

\bibitem{20} L.-M. Duan and G.-C. Guo, Phys. Rev. Lett, 79, 1953 (1997);

\bibitem{21} L.-M. Duan and G.-C. Guo, Phys. Rev. A, 57, 737 (1998).

\bibitem{22} P. Zanardi and M. Rasetti, Phys. Rev. Lett. 79, 3306 (1997).

\bibitem{23} P. Zanardi and M. Rasetti, Mod. Phys. Lett. B, 11, 1085 (1997).

\bibitem{24} D. A. Lidar, D. Bacon and K. B. Whaley, Phys. Rev. Lett. 81,
2594 (1998).

\bibitem{25} D. A. Lidar, D. Bacon, J. Kempe and K. B. Whaley, Phys. Rev. A,
63, 022306 (2001).

\bibitem{26} D. A. Lidar and K. B. Whaley, Springer Lecture Notes in Physics
622 (2003)

\bibitem{27} P. G. Kwiat, A. J. Berglund, J. B. Altepeter and A. G. White,
Science, 290, 498 (2000).

\bibitem{28} D. Kielpinski, V. Meyer, M. A. Rowe, C. A. Sackett, W. M.
Itano, C. Monroe and D. J. Wineland, Science, 291, 1013 (2001).

\bibitem{29} E. M. Fortunato, L. Viola, J. Hodges, G. Teklemariam and D. G.
Cory, New J. Phys 4:5 (2002).

\bibitem{30} L. Viola, E. M. Fortunato, M. A. Pravia, E. Knill, R. Laflamme,
D. G. Cory, Science, 293, 2059 (2001).

\bibitem{31} J. Wrachtrup and F. Jelezko, J. Phys: Condens. Matter, 16,
R1089 (2004).

\bibitem{32} F. Jelezko, T. Gaebel, I. Popa, A. Gruber and J. Wrachtrup,
Phys. Rev. Lett. 92, 076401 (2004).

\bibitem{33} M. V. Gurudev Dutt, L. Childress, L. Jiang, E. Togan, J. Maze,
F. Jelezko, A. S. Zibrov, P. R. Hemmer, M. D. Lukin, Science 316, 1312
(2007).

\bibitem{34} T. van der Sar, Z. H. Wang, M.S. Blok, H. Bernien, T. H.
Taminiau, D. M. Toyli, D. A. Lidar, D. D. Awschalom, R. Hanson and V. V.
Dobrovitski, Nature, 484, 82-86 (2012).

\bibitem{35} T. H. Taminiau, J. J. T. Wagenaar, T. van der Sar, F. Jelezko,
V. V. Dobrovitski and R. Hanson, Phys. Rev. Lett. 109, 137602 (2012).

\bibitem{36} A. Reiserer, N. Kalb, M. S. Blok, K. J. M. van Bemmelen, T. H.
Taminiau and R. Hanson, Phys. Rev. X 6, 021040 (2016).

\bibitem{36a} Y. Wang, M. Um, J.-H. Zhang, S.-M. An, M. Lyu, J.-N. Zhang, L.-M. Duan, D. Yum, K. Kim, arxiv/1701.04195 and refs therein.

\bibitem{36b} A. I. Lvovsky, W. Tittel, and B. Sanders, Nature Photon. 3,
706-714 (2009);

\bibitem{36c} M. Zhong,	M. P. Hedges, R. L. Ahlefeldt, J. G. Bartholomew, S. E. Beavan, S. M. Wittig, J. J. Longdell, and M. J. Sellars, Nature 517, 177--180 (2015).

\bibitem{36d}  Y. O. Dudin, L. Li, A. Kuzmich, Phys. Rev. A 87, 031801(R)
(2013).

\bibitem{37} F. Wang, C. Zu, L. He, W.-B. Wang, W.-G. Zhang, L.-M. Duan,
Phys. Rev. B 94, 064304 (2016).

\bibitem{38} L. Jiang, J. S. Hodges, J. R. Maze, P. Maurer, J. M. Taylor, D.
G. Cory, P. R. Hemmer, R. L. Walsworth, A. Yacoby, A. S. Zibrov, M. D.
Lukin, Science, 326, 267 (2009).

\bibitem{38a} P. Neumann, N. Mizuochi, F. Rempp, P. Hemmer, H. Watanabe, S. Yamasaki, V. Jacques, T. Gaebel, F. Jelezko, J. Wrachtrup, Science, 320, 1326 (2008).

\bibitem{38b} G. Waldherr, Y. Wang, S. Zaiser, M. Jamali, T. Schulte-Herbruggen, H. Abe, T. Ohshima, J. Isoya, J. F. Du, P. Neumann, J. Wrachtrup, Nature 506, 204(2014).

\bibitem{38c} Wolfgang Pfaff, Tim H. Taminiau, Lucio Robledo, Hannes Bernien, Matthew Markham, Daniel J. Twitchen, Ronald Hanson, Nature Phys. 9, 29 (2013).

\bibitem{38d} J. Cramer, N. Kalb, M. A. Rol, B. Hensen, M. S. Blok, M. Markham, D. J. Twitchen, R. Hanson, T. H. Taminiau, Nat. Commun. 7: 11526 (2016).

\bibitem{39} Supplementary materials.
\end{thebibliography}
\end{document}

\title{Supplementary Material "Room temperature entanglement storage
using decoherence free subspace in a solid-state spin system"}
\author{F. Wang$^1$, Y.-Y. Huang$^1$, Z.-Y. Zhang$^{1,2}$, C. Zu$^1$, P.-Y. Hou$^1$,
X.-X. Yuan$^1$, W.-B. Wang$^1$, W.-G. Zhang$^1$, L. He$^1$, X.-Y. Chang$^1$,
L.-M. Duan}
\affiliation{Center for Quantum Information, IIIS, Tsinghua University, Beijing 100084,
PR China}
\affiliation{Department of Physics, University of Michigan, Ann Arbor, Michigan 48109, USA}
\date{\today }

\begin{abstract}
In this supplementary material, we include experimental details and
numerical simulation for parameter calibration, quantum gates, and state
detection in weakly coupled nuclear spin systems. We also describe
experimental technique to control environment and realize general collective
noise model.
\end{abstract}

\date{\today }
\maketitle

\section{Experimental Setup}

The optical setup is similar to what was described in Ref \cite{1}. A major
difference is that another acoustic optical modulator (AOM) in double pass
configuration is added into the optical path right after the first AOM
double pass to constrain the leakage of green laser to a higher order.

The microwave signal is delivered into the system with a waveguide
transmission line fabricated on a cover glass, to which the diamond sample
is attached. The microwave field is generated by a carrier signal modulated
at an IQ mixer by two analog outputs of an Arbitrary Waveform Generator
(AWG) to control the relative phase of the signal. We add a switch
controlled by the digital output of the AWG after the IQ mixer to reduce the
influence of leakage of the carrier signal. The microwaves are then sent to
a high power amplifier and subsequently get delivered to the sample.

The radio frequency signal is generated by the analog channel of the AWG and
amplified through a high power amplifier. We fabricated a coplanar coil with
matching impedance on a PCB board to deliver radio frequency signal. To
achieve a reasonable nuclear spin Rabi frequency, the coil is detached from
the diamond surface with a distance about $2$ mm (The other surface is
attached to the cover glass).

A magnetic field of $B_{z}=480$ Gauss is applied along the NV symmetry axis
using a permanent magnet. The strong magnetic field is used to polarize the
intrinsic nitrogen spin and provide a relative strong Larmor frequency to
the nuclear spins compared to the nuclear spin hyperfine parameters. All the
experiments are performed at room temperature with $10^{6}$ repetitions for
measurement of each data point.

\begin{figure}[tbp]
\includegraphics[width=170mm]{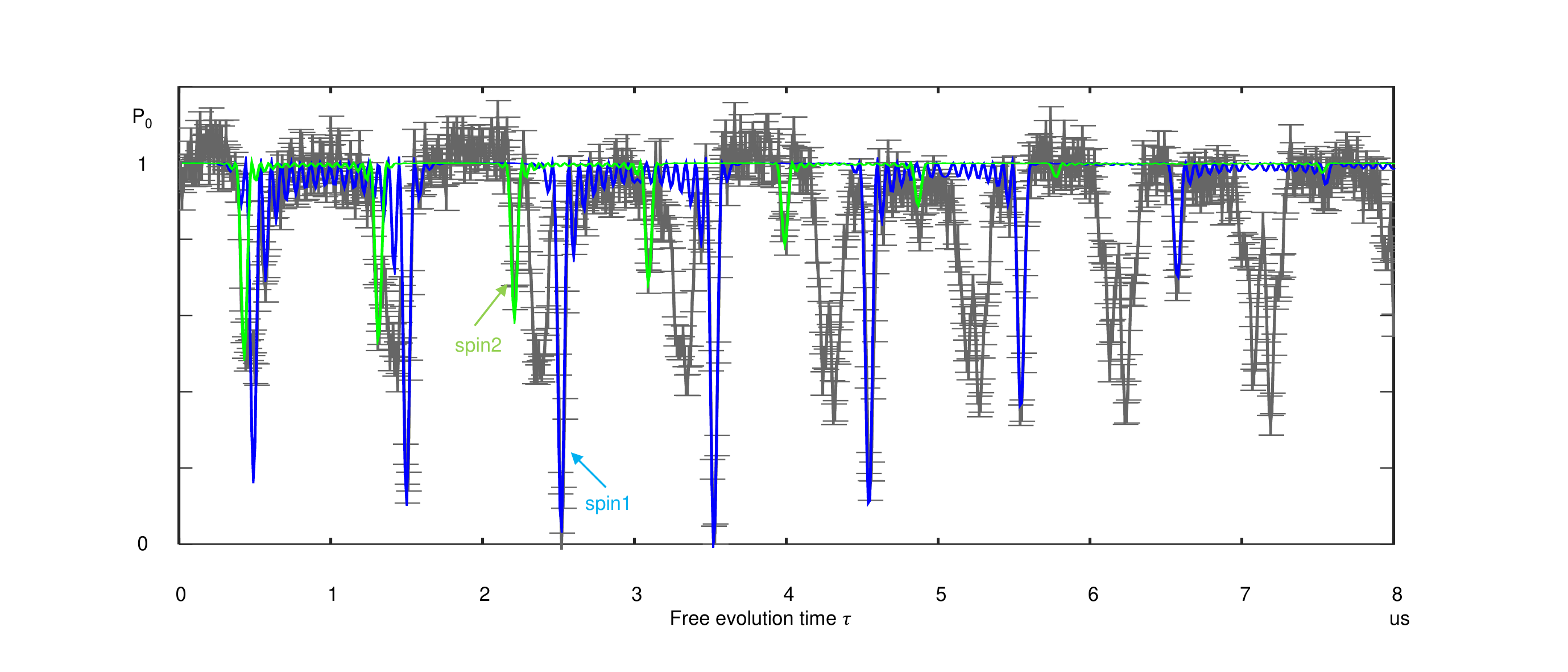}
\caption{Calibration of nuclear spin environment by the CPMG pulse sequence. What is shown is the measured coherence signal after
the CPMG decoupling sequence as a function of free evolution time $\protect\tau$. The
signal (grey) is taken with a magnetic field $B=490$ Gauss by preparing the
electronic spin in $(|0\rangle-i|1\rangle)/\protect\sqrt{2}$ state and
applying a XY8 sequence with $16$ $\protect\pi$ rotations before measuring
the electronic spin coherence by another $\protect\pi/2$ rotation around the
$-X$ axis. The free precession time $\protect\tau$ is taken with a step of $%
10ns$. Blue(green) is simulation with hyperfine parameters calibrated from
the nuclear spin ODMR signal for spin 1 (spin 2).}
\end{figure}

\begin{figure}[tbp]
\includegraphics[width=170mm]{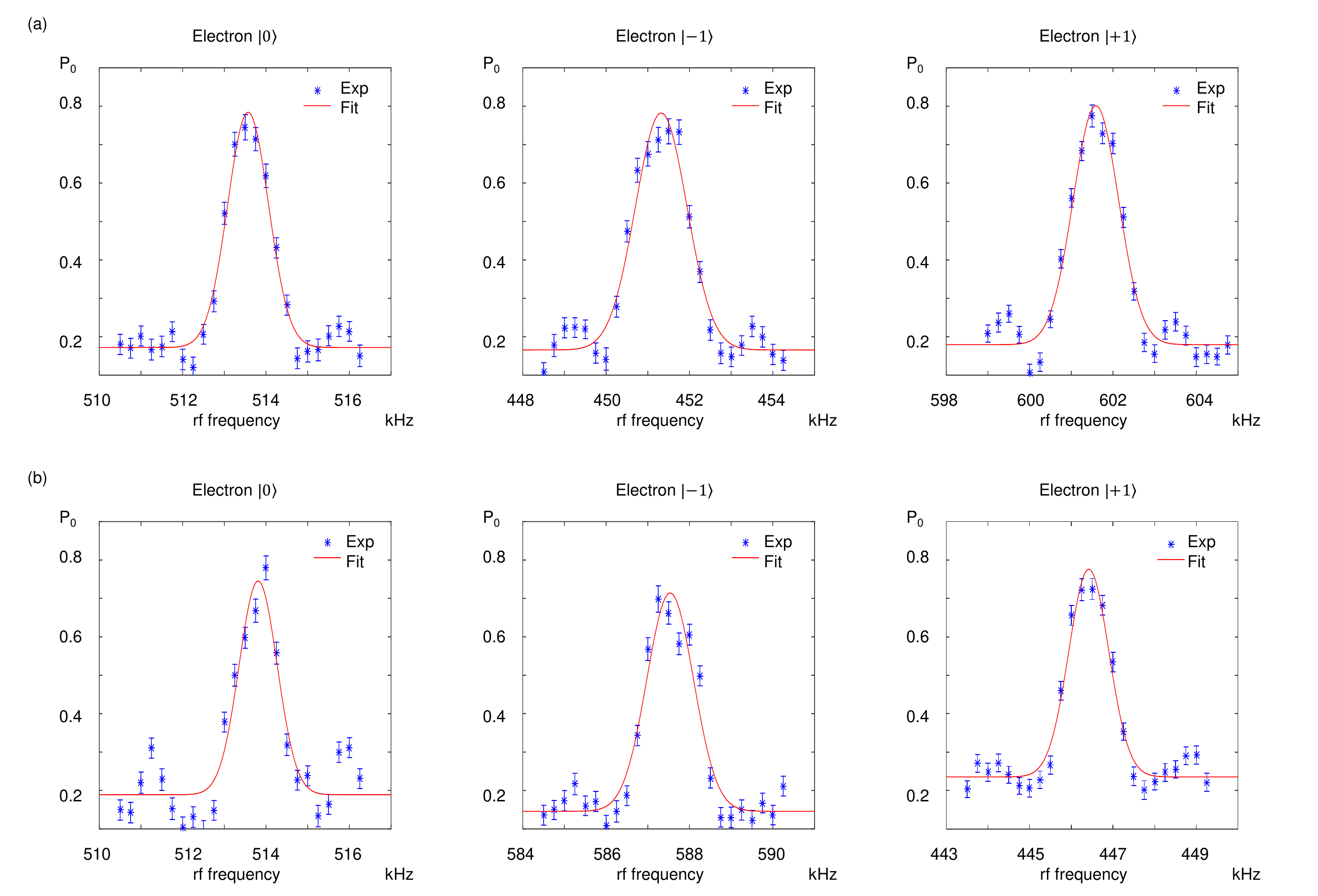}
\caption{Nuclear spin ODMR results for spin 1 and spin 2. The experimental scheme
is illustrated in Fig. 1 in the main text. (a,b) Resonant frequency with
electronic spin at $m_s=+1,0,-1$ state for nuclear spin 1 and 2.}
\end{figure}

\section{Calibration of nuclear-spin hyperfine interaction parameters}

Due to the anisotropic property of the hyperfine interaction, nuclear spins
undergo different evolutions with electronic spin at different eigenstates $%
m_{s}=+1,0,-1$. We consider the two-spin system composed of an electronic
spin with $m_{s}=0$ and $m_{s}=-1$, denoted as $|0\rangle $ and $|1\rangle $%
, and a nuclear spin with components of the spin operator denoted as $%
I_{x},I_{y},I_{z}$. With the electronic spin at $|0\rangle $ and $|1\rangle $
state, the nuclear spin Hamiltonian is denoted as $H_{0}$ and $H_{1}$,
respectively, which takes the form
\begin{equation}
H_{0}=\omega _{L}I_{z},
\end{equation}%
\begin{equation}
H_{1}=(\omega _{L}+A_{\parallel })I_{z}+A_{\perp }I_{x},
\end{equation}%
where $\omega _{L}$ is the nuclear Larmor frequency, $A_{\parallel }$ and $%
A_{\perp }$ are the parallel and transverse components of the hyperfine
parameters. Consider a simple equally-spaced sequence of $\pi $ rotations ($%
\tau -\pi -2\tau -\pi -\tau $) with pulse number $N=2$ and pulse interval
denoted by $2\tau $, the net result of this specific decoupling sequences is
that the nuclear spin rotates by an angle of $\phi $ around axis $\hat{n}_{0}
$ ($\hat{n}_{1}$) with the electronic spin at $m_{s}=0$ ($m_{s}=-1$) state.
When $\hat{n}_{0}\cdot \hat{n}_{1}=-1$, a resonance condition is satisfied
and the electronic spin gets entangled with the nuclear spin, thus the
electronic coherence collapses after the CPMG sequence (Fig. 1 grey). With $%
\omega _{L}\gg A_{\parallel },A_{\perp }$, the condition of resonance is
given by \cite{2}
\begin{equation}
\tau \approx \frac{(2k-1)\pi }{2\omega _{L}+A_{\parallel }}
\end{equation}%
where $k$ is the order of resonance. The probability that the coherence is
preserved is given by \cite{2}
\begin{eqnarray}
P &=&(M+1)/2, \\
M &=&1-(1-\hat{n}_{0}\cdot \hat{n}_{1})sin^{2}\frac{N\phi }{2}
\end{eqnarray}%
At resonance, with $\omega _{L}\gg A_{\parallel },A_{\perp }$, this equation
becomes \cite{2}
\begin{eqnarray}
P &=&(cos(Nm_{x})+1)/2, \\
m_{x} &=&\frac{A_{\perp }}{\sqrt{(A_{\parallel }+\omega _{L})^{2}+A_{\perp
}^{2}}}
\end{eqnarray}%
By fitting the experimental data on the measured electronic spin coherence
after the CPMG sequence to the numerical simulation of the corresponding
dynamics with the fitting parameters $A_{\parallel }$ and $A_{\perp }$,
single nuclear spins can be resolved to a resolution of about $10$ $kHz$
\cite{2}. As illustrated in the main text, to calibrate the nuclear spin
hyperfine parameters more precisely, we run nuclear spin ODMR experiments
with the electronic spin set at $m_{s}=+1,0,-1$ states (Fig. 2). With the
nucelar spin ODMR technique, the hyperfine parameters for the two nuclear
spins used in our experiment are determined with a high precision as
follows:
\begin{equation}
A_{\parallel 1}=-77.02(3)\text{ kHz},\text{ \ }A_{\perp 1}=114.5(1)\text{ kHz%
},
\end{equation}%
\begin{equation}
A_{\parallel 2}=71.03(3)\text{ kHz},\text{ \ }A_{\perp 2}=58.7(3)\text{ kHz},
\end{equation}%
where the number in the bracket denotes the standard deviation on the last
digit.

\begin{figure}[tbp]
\includegraphics[width=170mm]{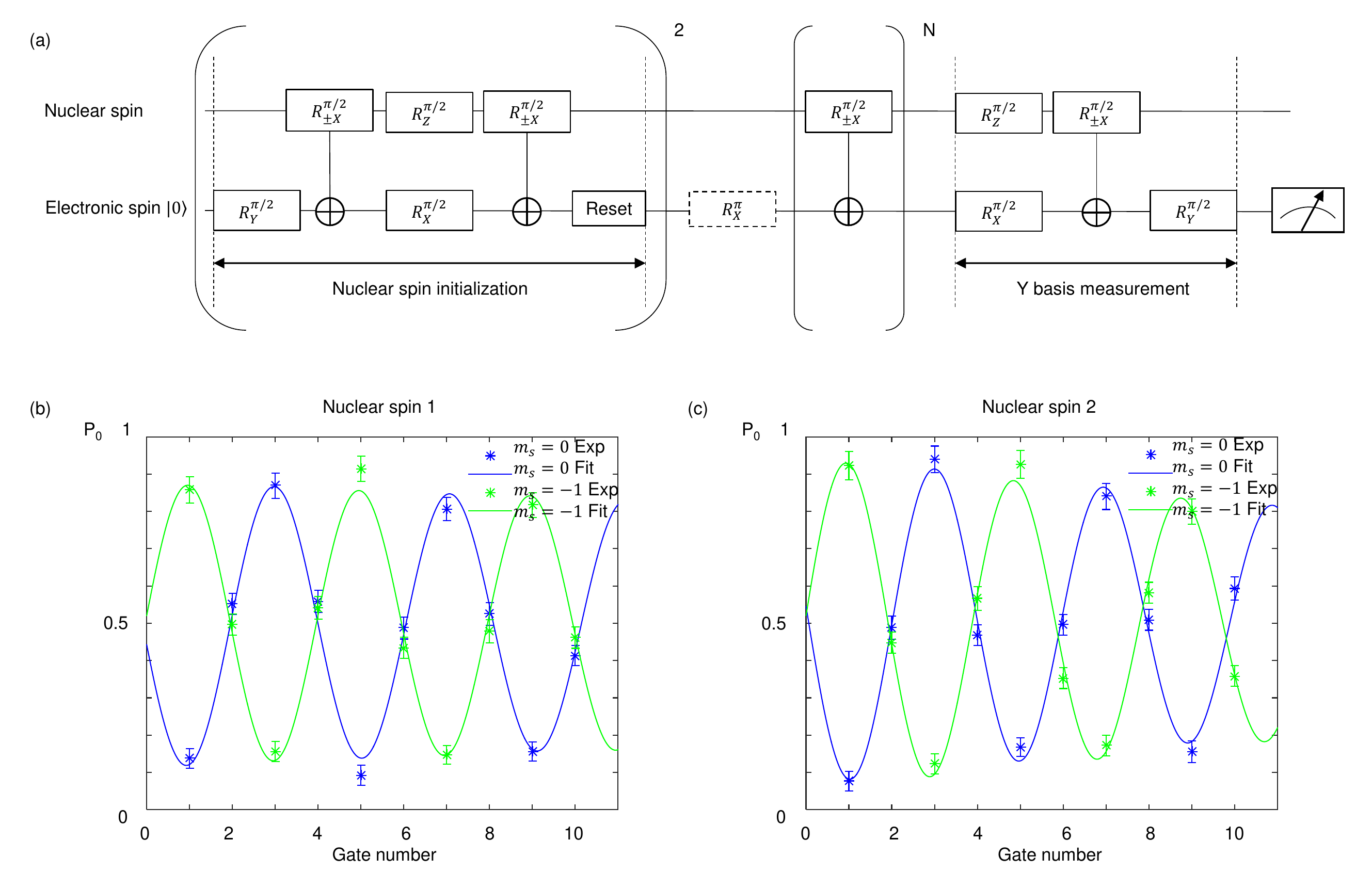}
\caption{Characterization of the conditional X gate for nuclear spin 1 and
2. (a) Experimental scheme. The nuclear spin is polarized by swapping
the electronic spin polarization onto the nuclear spin. An optional $\protect%
\pi$ rotation is applied to set the electronic spin to $m_s=-1$ state. After
that, the desired gate is applied on the nuclear spin for $N=1,...10$ times
with electronic spin at $m_s=0$ or $m_s=-1$ separately before measuring the
nuclear spin on the Y basis. (b,c) Characterization of conditional X gate
for nuclear spin 1 and 2. Solid lines are fit by the formula $sin(2\protect\pi N/4)*(1-bN)$ with $b=0.012$ and a standard deviation $\sigma=0.011$ in (b),
and with $b=0.025$ and a standard deviation $\sigma=0.014$ in (c).}
\end{figure}

\begin{figure}[tbp]
\includegraphics[width=170mm]{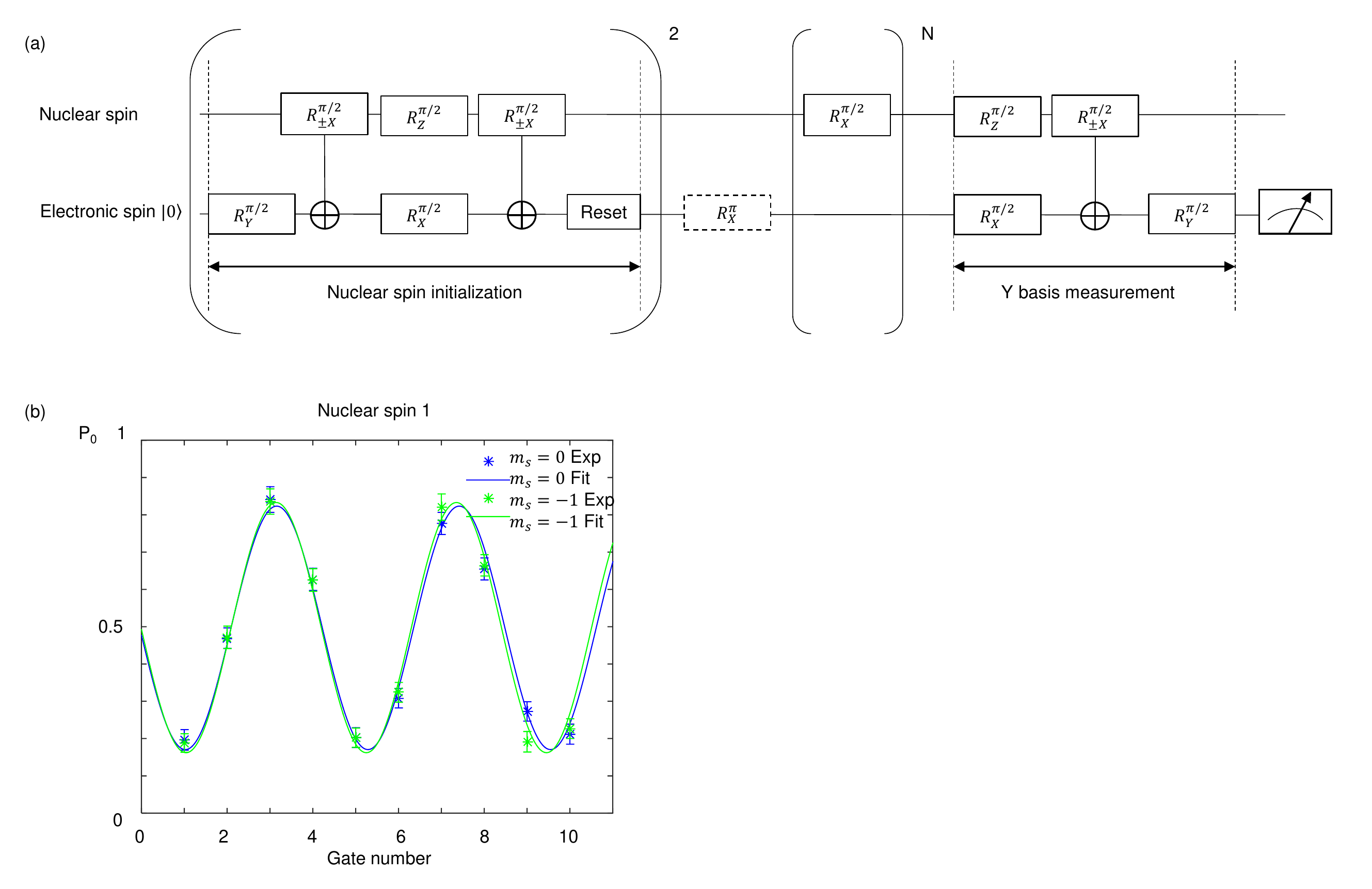}
\caption{Characterization of the unconditional X gate for nuclear spin 1. (a)
Experimental scheme. The nuclear spin is polarized by swapping the
electronic spin polarization onto the nuclear spin. An optional $\protect\pi$
rotation is applied to set the electronic spin to $m_s=-1$ state. After
that, the desired gate is applied on the nuclear spin for $N=1,...10$ times
with electronic spin at $m_s=0$ or $m_s=-1$ separately before measuring the
nuclear spin on the Y basis. (b) Characterization of unconditional X gate
for nuclear spin 1. Solid lines are fits by the formula $sin(2\protect\pi %
N/4)*(1-bN)$ with $b=0$ and a standard deviation $\sigma=0.013$.}
\end{figure}

\begin{figure}[tbp]
\includegraphics[width=170mm]{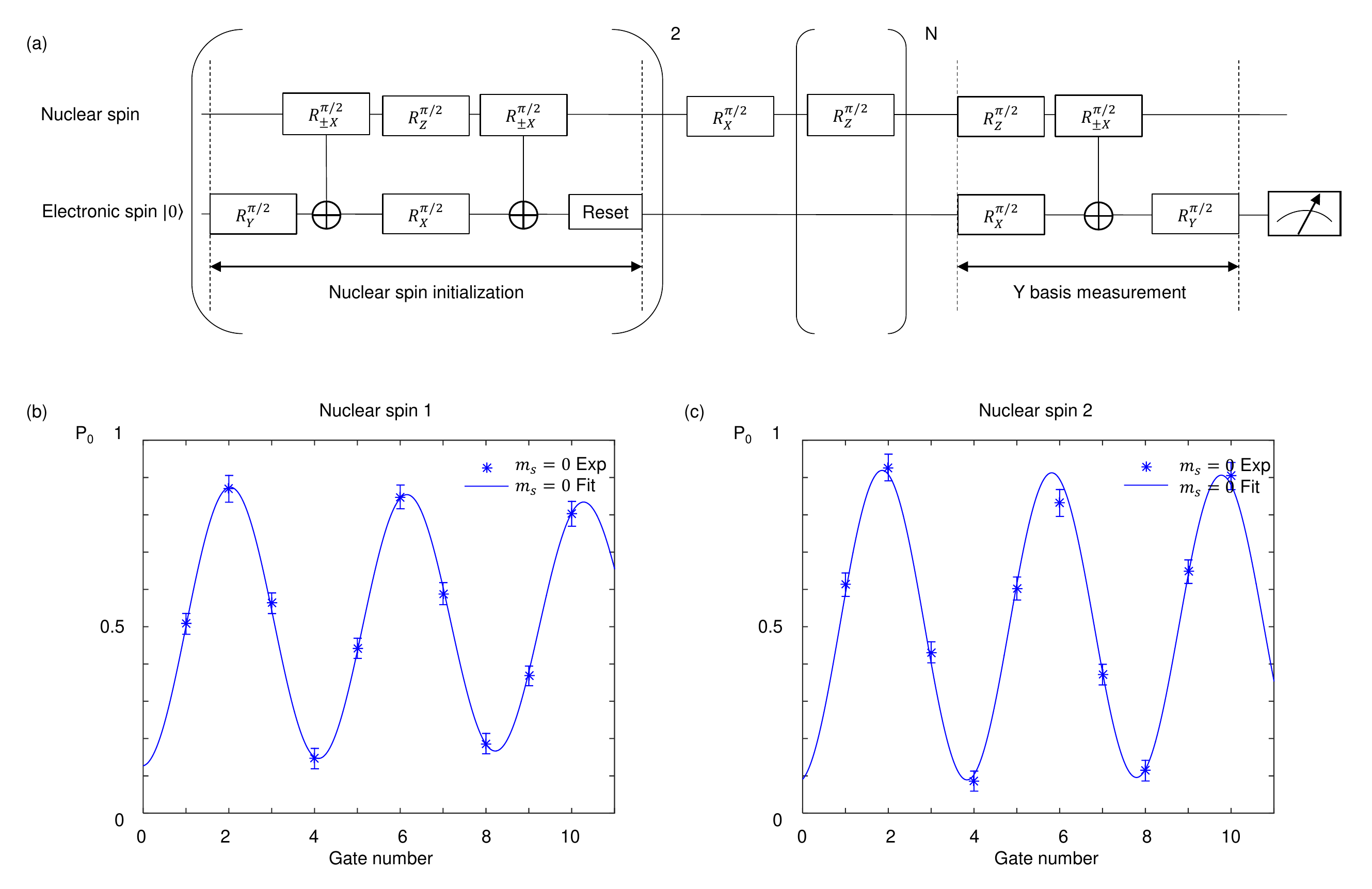}
\caption{Characterization of the unconditional Z gate for nuclear spin 1 and
spin 2. (a) Experimental scheme. The nuclear spin is polarized by swapping
the electronic spin polarization onto the nuclear spin. A $\protect\pi/2$
rotation is applied to set the nuclear spin to $(|0\rangle-i|1\rangle)/%
\protect\sqrt{2}$ state. After that, the desired gate is applied on the
nuclear spin for $N=1,...,10$ times with electronic spin at $m_s=0$ state
before measuring on the Y basis. (b,c) Characterization of unconditional Z
gate for nuclear spin 1 and spin 2. Solid lines are fits by the formula $sin(2\protect\pi%
N/4)*(1-bN)$ with $b=0.013$ and a standard deviation $\sigma=0.006$
in (b), and with $b=0.004$ and a standard deviation $\sigma=0.015$ in (c)}
\end{figure}

\section{Control of the two weakly coupled nuclear spins}

The conditional and unconditional gates are accomplished by the same pulse
sequence used to calibrate the nuclear spin hyperfine interaction
parameters. The parameters $\tau $ and $N$ (Table I) in each gate are
calculated using the calibrated hyperfine parameters. In Fig. 3, Fig. 4, and
Fig. 5, we apply the same gates (conditional X gate, unconditional X gate,
or unconditional Z gate) $N$ times (with $N=1,...10$) on the polarized
nuclear spin with the electronic spin initialized at different states.

\begin{table}[tbp]
\caption{Spin1, Spin2: gate parameters in the experiment}
\begin{center}
\begin{tabular}{l|c|c|c|c}
& $U$ & $\tau(\mu s)$ & $N$ & total time($\mu s$) \\ \hline
& $R_X^e(\pi/2)$ & $2.579$ & $7$ & 36.1 \\
spin1 & $R_X(\pi/2)$ & $4.123$ & $8$ & 66.0 \\
& $R_Z(\pi/2)$ & $0.047$ & $4$ & 0.4 \\ \hline
& $R_X^e(\pi/2)$ & $2.253$ & $19$ & 85.6 \\
spin2 & $R_Z(\pi/2)$ & $0.039$ & $4$ & 0.3 \\ \hline
\end{tabular}
\end{center}
\end{table}

The gate fidelity is mainly restricted by three factors: (i) The precision
of the hyperfine interaction parameters. (ii) The fluctuation of magnetic
field. (iii) The decoherence of the electronic spin during the pulse
sequence. To suppress the influence from the latter one, we choose two
nuclear spins with hyperfine parameters $A_{\parallel }$ around $\pm 50kHz$
so that they can be isolated from each other and the spin bath even at a
small pulse interval $\tau $. Therefore, the gate time can be controlled
within $100$ $\mu s$ and decoherence contribution by the electronic spin to
the gate infidelity is mitigated. From the slow decay of the oscillations in
Fig. 3, Fig. 4 and Fig. 5, we estimate a gate fidelity of $F\sim 0.988$ ($0.975
$) for the conditional X-gate on the nuclear spin 1 (2). The conditional
gate on the nuclear spin 2 has a lower fidelity as its pulse sequence has a
longer time and thus a larger contribution from the electronic spin
decoherence. The unconditional gate has a higher intrinsic fidelity: we do
not see noticeable fidelity decay after $10$ gates under experimental uncertainty, suggesting its
intrinsic fidelity $F>99\%$.

\section{Initialization and readout of single nuclear spins}

The single nuclear spin initialization and readout is obtained by measuring
the free evolution contrast of the nuclear spin coherence. In Fig. 6, the
nuclear spin is prepared to $(|0\rangle -i|1\rangle )/\sqrt{2}$ with the
electronic spin at $m_{s}=0$ ($m_{s}=-1$) for nuclear spin 1 (spin 2). We
characterize the nuclear spin coherence by projecting the nuclear spin phase
to the electronic spin population. We repeat the initialization process
twice and extract an initialization and readout fidelity of $F_{1}=0.896(6)$
and $F_{2}=0.873(9)$ for nuclear spin 1 and 2.

In experiments the initialization and readout fidelity of nuclear spins are always combined together as there is no direct ways to polarize or measure the state of nuclear spins separately. From the supplementary information of Ref. \cite{6}, the high initialization and readout fidelity indicates that the charge state initialization fidelity is either high ($>0.8$ from our experiment result) or not sensitive to the measurements. We briefly summarize the arguments here respectively under the following two assumptions (i) the re-initialization has no memory for the charge state or (ii) the re-initialization does not change the charge state.

Consider green laser initialization involves both spin states and charge states, the initial state of the electronic spin takes the form:
\begin{equation}
\rho_e=p_1\rho_0+p_2\rho_m+p_3\rho_s+p_4\rho_c
\end{equation}
where $\rho_0$ is the desired $m_s=0$ state, $\rho_m$ denotes the completely mixed state of $m_s=0$ and $m_s=-1$ states, $\rho_s$ denotes the other spin state $m_s=+1$, and $\rho_c$ denotes the $NV^0$ state. The probabilities satisfy the normalization $p_1+p_2+p_3+p_4=1$. In our normalized fluorescence contrast measurement, the $m_s=0$ state gives a signal of $1$, $m_s=\pm 1$ states and $NV^0$ give a signal of $0$.

Assume the unpolarized nuclear spin is in a completely mixed state $\rho_m$, the state of the initialized electronic spin and a single nuclear spin is:
\begin{equation}
\rho=\rho_e\otimes\rho_n=p_1(\rho_0\otimes\rho_m)+p_2(\rho_m\otimes\rho_m)+p_3(\rho_s\otimes\rho_m)+p_4(\rho_c\otimes\rho_m)
\end{equation}
After swapping the nuclear spin with the electronic spin
\begin{equation}
\rho=p_1(\rho_m\otimes\rho_0)+p_2(\rho_m\otimes\rho_m)+p_3(\rho_s\otimes\rho_m)+p_4(\rho_c\otimes\rho_m)
\end{equation}

Under the scenario that re-initialization of the electronic spin has no memory for the charge state, the state becomes:
\begin{equation}
\rho=(p_1\rho_0+p_2\rho_m+p_3\rho_s+p_4\rho_c)\otimes(p_1\rho_0+(1-p_1)\rho_m)
\end{equation}
Reading out the nuclear spin involves another swap gate between the nuclear and the electronic spin:
\begin{equation}
\rho=p_1(p_1\rho_0+(1-p_1)\rho_m)\otimes\rho_0+p_2(p_1\rho_0+(1-p_1)\rho_m)\otimes\rho_m+p_3\rho_s\otimes(p_1\rho_0+(1-p_1)\rho_m)+p_4\rho_c\otimes(p_1\rho_0+(1-p_1)\rho_m)
\end{equation}
Reading out electronic spin using green laser only yields non-zero signal for the electronic spin in the pure state $\rho_0$. Thus the maximum signal contrast is:
\begin{equation}
C_{max}=\frac{p_1^2+p_1p_2}{p_1}=p_1+p_2
\end{equation}
so that the maximum initialization and readout fidelity is calculated by:
\begin{equation}
F_{max}=\frac{1}{2}+\frac{C_{max}}{2}=\frac{1}{2}+\frac{p_1+p_2}{2}
\end{equation}
From the above equation and the initialization and readout fidelity measured in the experiment ($\sim 0.9$), we find $p_1+p_2\sim 0.8$, thus the charge state initialization fidelity $p_1+p_2+p_3>0.8$ under this scenario.

Alternatively, if the electron re-initialization does not change the charge state, the state after re-initialization becomes:
\begin{equation}
\rho=p_1(\frac{p_1\rho_0+p_2\rho_m+p_3\rho_s}{p_1+p_2+p_3}\otimes\rho_0)+(p_2+p_3)(\frac{p_1\rho_0+p_2\rho_m+p_3\rho_s}{p_1+p_2+p_3}\otimes\rho_m)+p_4\rho_c\otimes\rho_m
\end{equation}
After another swap gate,
\begin{equation}
\rho=p_1(\rho_0\otimes\frac{p_1\rho_0+p_2\rho_m}{p_1+p_2+p_3})+p_1(\frac{p_3\rho_s}{p_1+p_2+p_3}\otimes\rho_0)+(p_2+p_3)(\rho_m\otimes\frac{p_1\rho_0+p_2\rho_m}{p_1+p_2+p_3})+(p_2+p_3)(\frac{p_3\rho_s}{p_1+p_2+p_3}\otimes\rho_m)+p_4\rho_c\otimes\rho_m
\end{equation}
So that the maximum initialization and readout fidelity becomes:
\begin{equation}
F_{max}=\frac{1}{2}+\frac{p_1+p_2}{2(p_1+p_2+p_3)}
\end{equation}
This indicates that the initialization and readout fidelity is independent of the charge state initialization fidelity.

\begin{figure}[tbp]
\includegraphics[width=170mm]{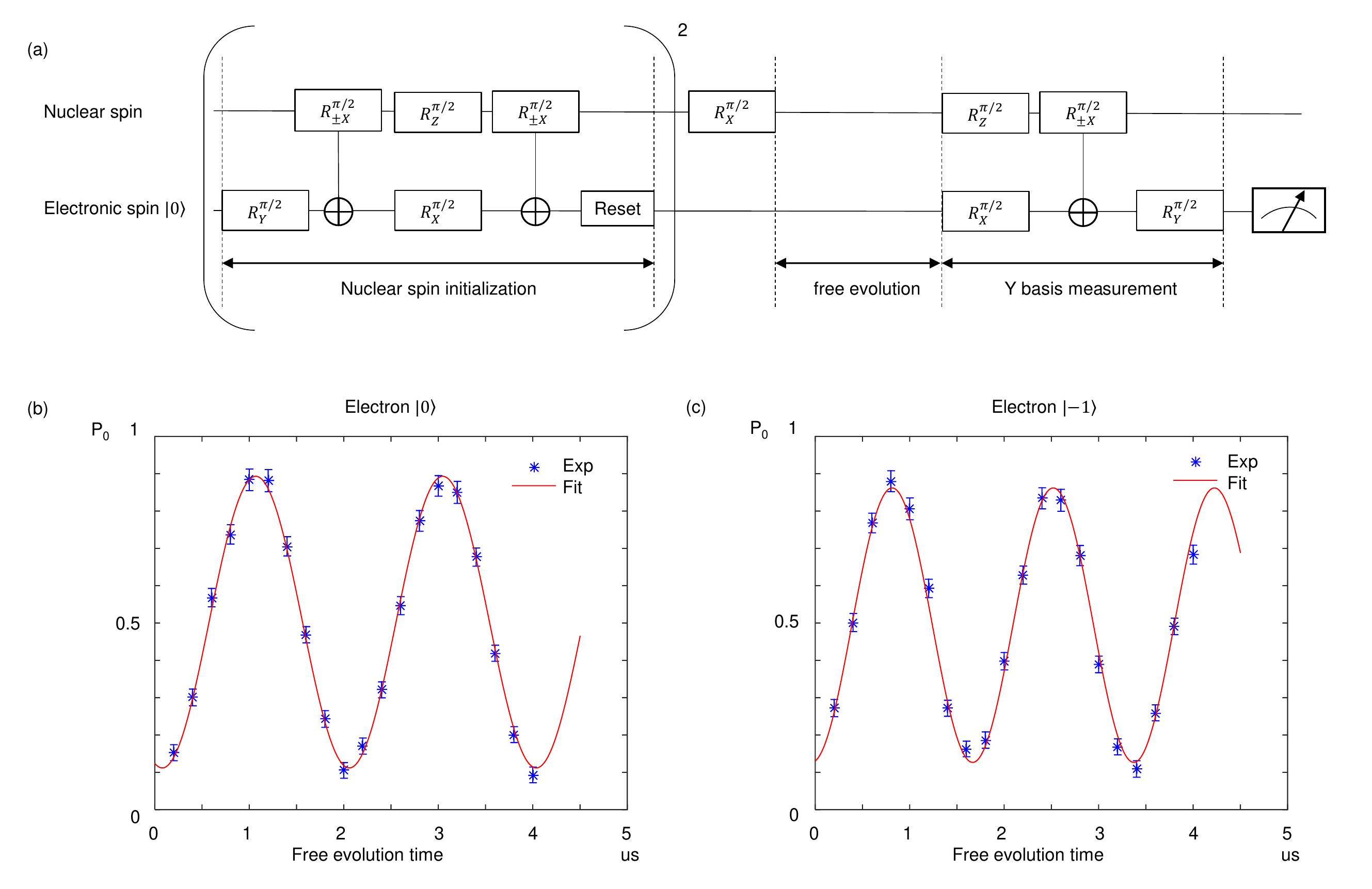}
\caption{Characterization of the initialization and readout fidelity for
nuclear spin 1 and spin 2. (a) Experimental scheme. The nuclear spin is
prepared to $(|0\rangle-i|1\rangle)/\protect\sqrt{2}$ state by a $\protect\pi%
/2$ rotation after the polarization. The measurement is performed by
swapping the nuclear coherence onto the electronic spin polarization. The
initialization and readout fidelity is obtained by measuring the contrast of
the evolution. (b,c) Free evolution for nuclear spin 1 and spin 2 as a
function of evolution time.}
\end{figure}

\section{Two-bit quantum state tomography}

The two-bit tomography consists of three-basis (X, Y, Z) single-bit
measurements on the two nuclear spins separately and nine two-bit
correlation measurements. All the measurements are performed by mapping the
nuclear spin information onto the electronic spin population. Figure 7 shows
our experimental scheme for the single-bit and two-bit measurements. The
nuclear spin density matrix is extracted from the two-bit tomography result
with a maximum likelihood calculation \cite{3}.

\begin{figure}[tbp]
\includegraphics[width=180mm]{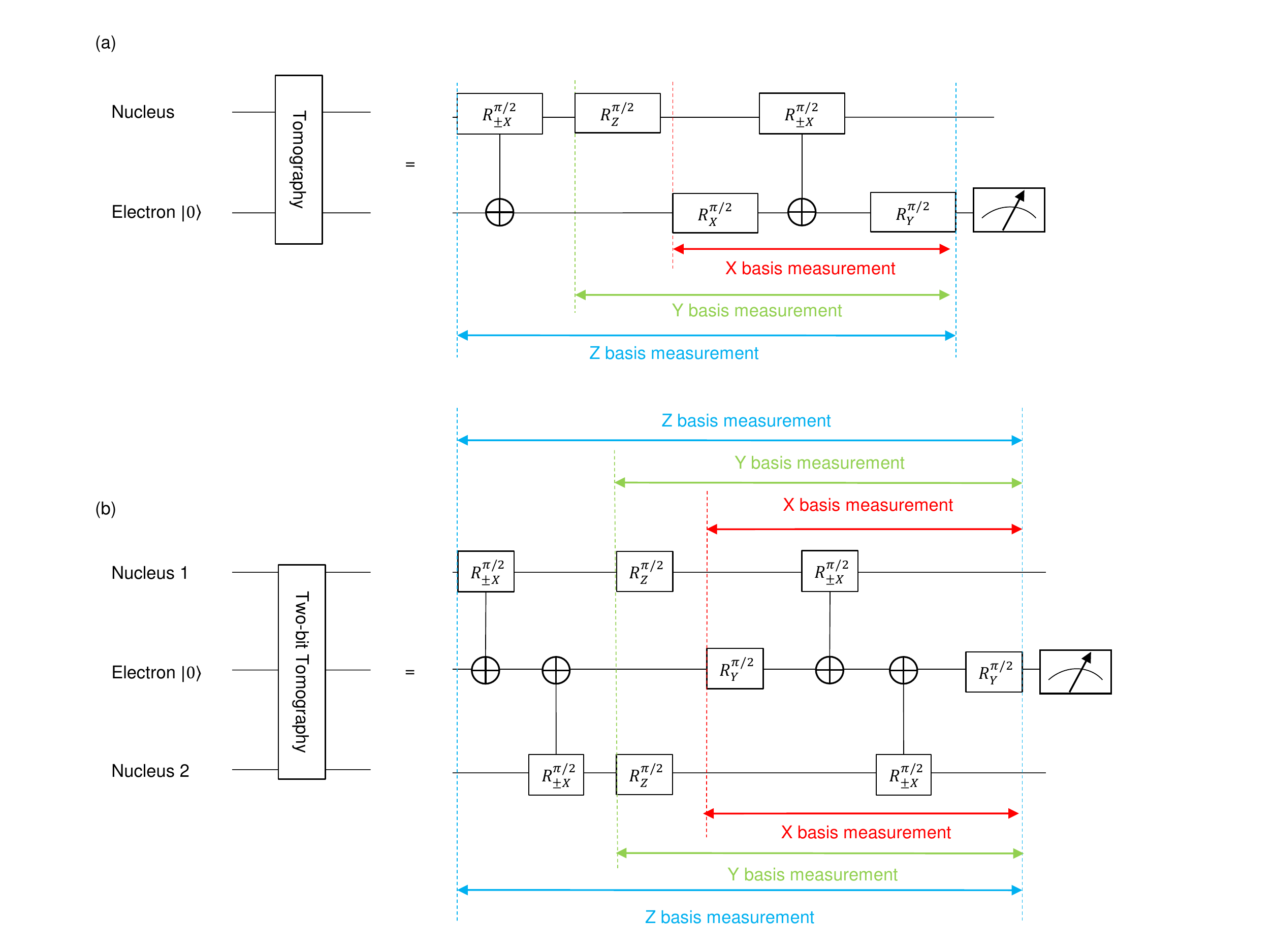}
\caption{Quantum state tomography scheme. (a) Single-bit tomography scheme of the three
bases on the two nuclear spins separately. (b) Correlation measurements of
the $9$ bases on the two nuclear spins cooperatively. }
\end{figure}

\section{Numerical simulation}

The numerical simulation is performed in the rotating frame in the
three-qubit system composed of electronic spin and the two target nuclear
spins. We assume a noiseless environment without decoherence and relaxation.
The $350$ ns green laser pumping is simulated by an instant reset of the
electronic spin. After the reset, the nuclear spin state is given by the
partial trace over the electronic spin state. All the parameters in the
simulation is the same as those calibrated by the experiments.

\section{Crosstalk between the two nuclear spins}

As the nuclear hyperfine interactions are always on, gates on one nuclear
spin will affect other nuclear spins and lead to unwanted operations. There
are two types of possible operations on other nuclear spins: (i) Conditional
or unconditional X operations. (ii) Z rotations.

Because conditional and unconditional X operations only happen at specific
time $\tau $ with a very narrow bandwidth, the first type of influence can
be avoided by choosing $\tau $ to bypass X rotations on other nuclear spins.
To reduce the influence from the unwanted Z rotation, we track the phase of
each nuclear spin in experiment through simulation and compensate
accumulated phase at proper time in experimental sequence using the right
parameters fixed from the simulation.

\section{Realization of general collective noise model}

By injecting radio frequency (rf) noise into the system to drive the nuclear
spin transitions, we can realize any collective noise model. The rf signal
is centered at the nuclear spin Larmor frequency with a bandwidth of $10$
kHz. To model a noisy environment with time correlation function of the
shape $exp(-R|\tau |)$, we add up all the frequency components weighted with
function $\sqrt{\frac{2\delta \omega R}{(2\pi n\Delta \omega )^{2}+R^{2}}}$,
where $\Delta \omega =1$ kHz is the discretization step. The noise is turned
on $5$ $\mu $s after the entanglement preparation step and turned off $5$ $%
\mu $s before the state tomography measurement to avoid the ac Stark shift
on the electronic spin caused by the rf signal.

\section{Influence of the magnetic field fluctuation}

The magnetic field is calibrated by measuring electronic ODMR signal every
two hours during the experiments. Due to the fluctuation in the lab
temperature, the magnetic field fluctuates on the order of $0.2$ G. The
fluctuation leads to gate errors accumulated on nuclear spin 1 as well as
unwanted phase evolution on nuclear spin 2 in the entangling process. The
induced phase fluctuation over the $10^{6}$ repetitions of measurements of
each experimental density matrix element leads to a drop of the measured
entanglement fidelity. In our numerical simulation, we find that a magnetic
field fluctuation with a Gaussian shape and a standard deviation of $0.15$ G
leads to dropping of the entanglement fidelity from $1$ to $0.92$.